\documentclass[journal=jacsat,manuscript=article]{achemso}

\usepackage[version=3]{mhchem} 
\usepackage{color,soul,enumerate}
\usepackage{multirow}
\usepackage[table]{xcolor}
\usepackage{comment}
\usepackage{booktabs}
\usepackage{graphicx}

\title{Atomistic simulations of the crystalline-to-amorphous transformation of $\gamma$-\ce{Al2O3} nanoparticles: delicate interplay between lattice distortions, stresses, and space charges}

\author{Simon Gramatte}

\affiliation[inst1]{Laboratory for Advanced Materials Processing, Empa - Swiss Federal Laboratories for Materials Science and Technology, Feuerwerkerstrasse 39, 3602 Thun, Switzerland}

\alsoaffiliation[inst2]{Laboratory for Joining Technologies and Corrosion, Empa - Swiss Federal Laboratories for Materials Science and Technology, Ueberlandstrasse 129, 8600 Duebendorf, Switzerland}

\alsoaffiliation[inst3]{Laboratoire Interdisciplinaire Carnot de Bourgogne, UMR 6303 CNRS-Université Bourgogne Franche-Comté, 9 Avenue A. Savary, Dijon Cedex, France}

\author{Lars P.H. Jeurgens}
\affiliation[inst2]{Laboratory for Joining Technologies and Corrosion, Empa - Swiss Federal Laboratories for Materials Science and Technology, Ueberlandstrasse 129, 8600 Duebendorf, Switzerland}
\author{Olivier Politano}
\affiliation[inst3]{Laboratoire Interdisciplinaire Carnot de Bourgogne, UMR 6303 CNRS-Université Bourgogne Franche-Comté, 9 Avenue A. Savary, Dijon, France}
\author{Jose Antonio Simon Greminger}
\affiliation[inst1]{Laboratory for Advanced Materials Processing, Empa - Swiss Federal Laboratories for Materials Science and Technology, Feuerwerkerstrasse 39, 3602 Thun, Switzerland}
\author{Florence Baras}
\affiliation{Laboratoire Interdisciplinaire Carnot de Bourgogne, UMR 6303 CNRS-Université Bourgogne Franche-Comté, 9 Avenue A. Savary, Dijon, France}
\author{Angelos Xomalis}
\affiliation{Laboratory for Mechanics of Materials \& Nanostructures, Empa - Swiss Federal Laboratories for Materials Science and Technology, Feuerwerkerstrasse 39, 3602 Thun, Switzerland}
\author{Vladyslav Turlo}
\affiliation[inst1]{Laboratory for Advanced Materials Processing, Empa - Swiss Federal Laboratories for Materials Science and Technology, Feuerwerkerstrasse 39, 3602 Thun, Switzerland}
\email{vladyslav.turlo@empa.ch}
\phone{+41 58 765 6318}

\keywords{COMB3 molecular dynamics, alumina nanoparticles, amorphization, thermal stability, local coordination sphere}
\begin{document}

\begin{abstract}

The size-dependent phase stability of $\gamma$-\ce{Al2O3} was studied by large-scale molecular dynamics simulations over a wide temperature range from 300 to 900 K. For the $\gamma$-\ce{Al2O3} crystal, a bulk transformation to $\alpha$-\ce{Al2O3} by an FCC-to-HCP transition of the O sublattice is still kinetically hindered at 900 K. However, local distortions of the FCC O-sublattice by the formation of quasi-octahedral Al local coordination spheres become thermally activated, as driven by the partial covalency of the Al-O bond.
On the contrary, spherical $\gamma$-\ce{Al2O3} NPs (with sizes of 6 and 10 nm) undergo a crystalline-to-amorphous transformation at 900 K, which starts at the reconstructed surface and propagates into the core through collective displacements of anions and cations, resulting in the formation of $7$- and $8$-fold local coordination spheres of Al. In parallel, the reconstructed Al-enriched surface is separated from the stoichiometric core by a diffuse Al-depleted transition region. This compositional heterogeneity creates a disbalance of charges inside the NP, which induces a net attractive Coulombic force that is strong enough to reverse the initial stress state in the NP core from compressive to tensile. These findings disclose the delicate interplay between lattice distortions, stresses, and space-charge regions in oxide nanosystems. A fundamental explanation for the reported expansion of metal-oxide NPs with decreasing size is provided, which has significant implications for e.g. heterogeneous catalysis, NP sintering, and additive manufacturing of NP-reinforced metal matrix composites.
\end{abstract}

\section{Labels - Keywords}
Keywords: COMB3 molecular dynamics, alumina nanoparticles, amorphization, thermal stability, surface stress



\section{Introduction}
\label{sec:Intro}
Aluminum oxide (\ce{Al2O3}) exists in a wide variety of amorphous and crystalline polymorphs with different functional properties, strongly dependent on the synthesis, processing and operating conditions \cite{Wilson1980, Levin1998,Kovarik2021}. According to bulk equilibrium thermodynamics, $\alpha$-\ce{Al2O3}, also called corundum, is the only stable bulk \ce{Al2O3} polymorph \cite{Wriedt1985}, which implies that all other \ce{Al2O3} polymorphs are metastable as a (strain-free) bulk phase. However, in accordance with Ostwald's Rule of Stages \cite{Ostwald1897}, an experimentally-observed alumina phase does not necessarily correspond to the bulk-preferred equilibrium phase but can be a transition phase that is closest to the stability of its parent state. For example, the calcination of aluminum oxyhydroxide ($\gamma$-\ce{AlOOH}, boehmite) to $\alpha$-\ce{Al2O3} becomes thermally activated at around 800 K and is associated with a polymorphic phase transformation sequence from $\gamma\rightarrow\delta\rightarrow\theta\rightarrow\alpha$-\ce{Al2O3}.\cite{Wilson1980,Levin1998,Zhou2019, Kovarik2021} The transition alumina phases typically have a face-centered cubic (FCC) arrangement of oxygen atoms, while corundum has a hexagonal close-packed (HCP) O-sublattice \cite{Wilson1980,Levin1998,Zhou2019, Kovarik2021}.
The structural differences between the FCC-type alumina transition phases arise from the different distributions of the \ce{Al} cations and their vacancies over the tetrahedral and octahedral interstices of the (partially distorted) FCC O-sublattice, which strongly depends on the synthesis pathway.\cite{Levin1998,Kovarik2014,Kovarik2015,Zhou2019,Aboulkhair2019,Kovarik2021}

As demonstrated by theoretical investigations \cite{Jeurgens2000} and verified experimentally \cite{Reichel2008a}, the amorphous \ce{Al2O3} polymorph (\textit{am}-\ce{Al2O3}) can be thermodynamically preferred in the nanoscale range as long as the higher bulk Gibbs energy of the amorphous phase is overcompensated by its lower sum of surface and interfacial energies \cite{Jeurgens2009}. Similarly, other metastable \ce{Al2O3} polymorphs, including $\gamma$-\ce{Al2O3}, can be thermodynamically stable with respect to $\alpha$-\ce{Al2O3} due to the lower sum of excess Gibbs energies associated with surfaces and/or interfaces, which can include surface and/or interface stress contributions \cite{McHale1997a,Mavric2019,Reichel2007}. For example, at ambient temperatures and pressures, $\gamma$-\ce{Al2O3} has a lower surface energy than $\alpha$-\ce{Al2O3}, thus stabilizing the $\gamma$-\ce{Al2O3} phase for NP sizes below 12 nm \cite{McHale1997, Laurens2020}. Since the surface energy of an amorphous oxide phase is generally lower than that of the respective crystalline oxide polymorph \cite{Reichel2008b}, crystalline \ce{Al2O3} NPs can be surrounded by an amorphous shell (depending on size, temperature and pressure), as also commonly observed for surficial oxide films.\cite{Jeurgens2000, Luo2008, Reichel2008b} For \ce{Al2O3} NP sizes below 4 nm, the entire NP can be stabilized in an amorphous state.\cite{Tavakoli2013}. Notably, commercially available spherically-shaped \ce{Al2O3} NPs, as prepared by, e.g. flame spray pyrolysis, generally have an average primary particle size in the range of 6 - 35 nm \cite{Laine2006, Tok2006, Okonkwo2022}.

Stabilization of metastable \ce{Al2O3} polymorphs by surfaces, interfaces, and/or residual stresses has a large impact on key technologies, such as microelectronics, corrosion resistance, advanced manufacturing, and catalysis. For example, \textit{am}-\ce{Al2O3} is utilized as a barrier-type film in dielectric capacitors \cite{Ambrosio2017}, while \textit{am}-\ce{Al2O3} NPs have been shown to be an effective catalyst for biodiesel conversion \cite{Amini2013} or fluoride removal in water treatment \cite{Kang2015}. Moreover, \ce{Al2O3} NPs are among the most common materials for oxide-dispersed strengthening of metals and alloys, as produced by additive manufacturing (3D printing) techniques \cite{Redsten1995,Naser1997, Buchbinder2011, Sing2016, Guo2018, Leparoux2018}. Dispersed \ce{Al2O3} NPs act as nucleation sites during rapid cooling and solidification of the melt, resulting in grain refinement \cite{Buchbinder2011, Sing2016, Guo2018}. Furthermore, the strong interaction and pinning of lattice dislocations with ceramic NPs lead to Orowan strengthening of the material,\cite{Zhang2006} while Zener pinning of grain boundaries ensures improved thermal stability of the NP-reinforced metal matrix composite over a wide range of temperatures\cite{Manohar1998}.

Surprisingly, despite the broad use of \ce{Al2O3} NPs, still limited knowledge exists on the defect structure, phase stability and chemical reactivity as function of e.g. NP size and temperature \cite{Knozinger1987,Prins2020,Kovarik2015,Busca2014}. This might be rationalized by the fact that such fundamental knowledge requires tedious analysis of structural defects and associated local coordination spheres of the constituent ions (i.e. \ce{Al} cations and \ce{O} anions) in dependence of the synthesis, processing, and operating conditions. Atomistic simulations provide a powerful tool for such atomistic-scale analysis of NPs and their surfaces over a wide temperature range. Density functional theory (DFT) calculations of the structure of $\gamma$-\ce{AlO3} have mainly focused on the distribution of the \ce{Al} cations and their vacancies in the interstitial sites of the FCC \ce{O}-sublattice \cite{Gutierrez2001,Paglia2005,Pinto2004}, indicating that \ce{Al} cations can populate octahedral and tetrahedral interstices in the \ce{O}-sublattice at 0 K which is consistent with experimental studies at room temperature(RT)\cite{Prins2020, Stuart2021,Ferreira2011,Ayoola2020}. However, atomistic modeling of $\gamma$-\ce{Al2O3} at finite temperatures requires switching from DFT calculations to Molecular Dynamics (MD) simulations. On modern supercomputers, MD simulations can treat nanoscale structures that are composed of millions or even billions of atoms\cite{Zepeda-Ruiz2017,Shibuta2017,Zepeda-Ruiz2021}; however, the MD time scale is limited to nanoseconds due to the high computational cost of the interatomic potential and the inability to parallelize the time for complex processes \cite{Politano2015,Turlo2017a}. The interatomic potentials are parameterized using either experimental data or {\it ab initio} data \cite{Brazdova2013}. Several empirical interatomic potentials for \ce{Al2O3} have been developed, such as those by Alvarez \textit{et al.} \cite{Alvarez1994,Alvarez1995}, Vashishta \textit{et al.} \cite{Vashishta2008, Vashishta1990}, Woodley \cite{Woodley2011} and Streitz and Mintmire \cite{Streitz1994}. However, their applicability for predicting the structure of \ce{Al2O3} NPs has only been compared at RT \cite{Laurens2020}. The potentials by Alvarez \textit{et al.} and Streitz and Mintmire \cite{Streitz1994} both properly predict the experimentally observed, size-dependent polymorphic phase transformations in \ce{Al2O3} NPs at RT \cite{McHale1997, Tavakoli2013}. This despite the fact that the potential by Alvarez \textit{et al.} only accounts for Coulombic forces and steric repulsion (i.e. it does not account for variable charges). The potential by Streitz and Mintmire \cite{Streitz1994} is tailored to describe more complex \ce{Al}-\ce{Al2O3} interactions, but the angular contributions to the interatomic forces originating from the covalent character of the Al-O bond are not accounted for. Only the Second-Moment tight-binding potential by Salles \textit{et al.} explicitly takes into account the mixed ionic–covalent character of the Al-O bond but has only been applied to predict the relative phase stability of crystalline alumina polymorphs,\cite{Salles2016} thus neglecting the possible stabilization of amorphous polymorphs in the nano-scale range.

To better bridge the gap between {\it ab initio} and empirical potentials, reactive interatomic potentials, such as Charge Optimized Many-Body (COMB) and ReaxFF potentials, have been developed by merging variable charge electrostatic interactions with empirical potentials, thus representing different chemical bonding modes, Van-der-Waals forces, as well as quantum size effects \cite{Liang2013a,Liang2013b, VanDuin2001}. Reactive potentials determine the charges on every atom with a charge equilibration (QeQ) method that minimizes the electrostatic potential energy of the system by applying a charge-neutrality constraint. Therefore, reactive potential-based molecular dynamics simulations can describe different types of bonding (covalent, ionic, metallic) independently, which allows simulations of more complex systems and chemistry on longer time and length scales than in {\it ab initio} MD \cite{Slapikas2020,Senftle2016}. In the case of alumina, the third-generation COMB (COMB3) potential is performing much better than ReaxFF in terms of internal stress development in nanoscale systems with free surfaces and interfaces, in line with experimental observations \cite{Luu2022}. Moreover, the COMB3 potential is fully optimized for the Ni-Al-O system \cite{Kumar2015}, allowing for future studies of high-temperature interfaces between Ni-Al alloys/intermetallic compounds and alumina NPs.

Therefore, we applied the COMB3 potential to predict the phase stability of the bulk $\gamma$ -\ce{Al2O3} as well as of $\gamma$-\ce{Al2O3} NPs over a wide temperature range from RT to 2700 K (i.e. above the bulk melting point of alumina). The MD predictions are compared with the wealth of experimental and theoretical studies on the phase transformation sequence of bulk $\gamma$-\ce{Al2O3} to $\alpha$-\ce{Al2O3} with increasing temperature. Particular attention is paid to temperature- and surface-induced local distortions of the FCC O-sublattice, as accompanied by rearrangements of \ce{Al} atoms over (distorted) octahedral and/or tetrahedral interstices, as well as the development and stabilization of heterogeneous stress states during NP amorphization, which has major implications for understanding metal oxide nanoparticles. \

\section{Materials and Methods}

MD simulations were carried out with the Large-scale Atomic/Molecular Massively Parallel Simulator (LAMMPS) \cite{Plimpton1995a,Thompson2022} using the COMB3 potential for \ce{Al}-\ce{O} systems \cite{Choudhary2015a}. While being the most versatile classical potential, COMB3 suffers from low performance due to the charge equilibration procedure and small timesteps, which puts strong limitations on the time- and length-scales that can be resolved with such potentials. Our tests showed that the best compromise between computing time and accuracy of the charge equilibration (Qeq) procedure during thermal equilibration is achieved by executing it every 100 MD time steps with a precision (i.e. a convergence criterion for charge equilibration) of 0.01, and every 5 MD time steps with a precision equal to 0.05 (as also applied in Ref. \cite{Slapikas2020}). The equations of motion were integrated with a time step of 0.1 fs, which is one order of magnitude smaller than the usual timestep for classical potentials (1 fs) \cite{Slapikas2020, Arifin2019}. To ensure the thermal equilibrium in a canonical ensemble (NVT) \cite{Fan2010}, a Langevin thermostat \cite{Schneider1978}, commonly used in COMB3 simulations \cite{Slapikas2020}, was applied  with temperature damping parameters of 0.1 fs for \ce{Al} atoms and 0.05 fs for \ce{O} atoms. By applying atom-mass-dependent damping parameters for the \ce{O} and \ce{Al} atoms, the atom velocity is balanced according to mass-specific frictional drag, making the simulation more stable, particularly for the high-temperature runs. Each sample was initially pre-equilibrated for 5 ps until the velocity distribution of the particles followed a Maxwell-Boltzmann distribution \cite{Haile1992}, and was used as the starting configuration for further MD simulations.

The experimentally validated structure of $\gamma-$\ce{Al2O3} with randomly dispersed Al cation vacancies over the tetrahedral interstitial sites in the FCC O sublattice with a bulk lattice parameter of $a_0 = 7.9 \text{\AA}$ \cite{Villars2016} was used in this work. First, a bulk $\gamma-$\ce{Al2O3} crystal with dimensions of 5.5$\times$5.5$\times$5.5 $\:\text{nm}^3$ containing 18302 atoms was used to access the thermal stability and structure of $\gamma$-\ce{Al2O3} at 900 K (about 100 K below the bulk transformation temperature towards $\alpha$-\ce{Al2O3}). The atomic positions after structure relaxation were transformed into virtual diffraction patterns that can be compared to the corresponding experimental patterns \cite {Coleman2015, Coleman2016}. Next, a larger simulation box with dimensions of 20$\times$20$\times$20$\:\text{nm}^3$ was used to create spherical $\gamma-$\ce{Al2O3} NPs by filling the corresponding spherical region in the center of the box with atoms. As verified in the present study, the stoichiometric composition with overall charge neutrality is preserved for NP diameters $d>5 $ nm. The $\gamma-$\ce{Al2O3} NPs with $d = 6$ nm and $d = 10$ nm chosen in this work contained 16765 and 56578 atoms, respectively. A NP diameter of $d = 6$ nm corresponds to the lowest average NP size accessible to experimentalists \cite{Laine2006, Tok2006, Okonkwo2022}, whereas $d = 10$ nm is the largest size for which a reasonable time scale can be resolved with the COMB3 potential. Finally, the thermal stability of $\gamma$-\ce{Al2O3} NPs at 300 K, 900 K, and 2700 K (i.e. above the bulk melting point of alumina) was investigated, and the resulting structural transitions were followed by accessing energy per atom, local atomic displacements, and local hydrostatic stress as given by LAMMPS.

The Open VIsualization TOol (OVITO) was used for data visualization and post-processing analysis \cite{Stukowski2010b}. The Radial Pair Distribution functions, $g(r)$, were calculated with a cutoff radius of 10\AA. The standard $g(r)$ implemented in OVITO is only valid for simulation cells with periodic boundary conditions. To avoid the effects of the free surface in the $g(r)$ analysis of the NPs, only the atoms with a distance of 1 nm or more from the free surface were considered \cite{Kopera2018}. Nearest neighbor analysis (NNA) of the O sublattice was performed by \textit{Ackland and Jones} Analysis (AJA) \cite{Ackland2006}, which combines spatial and angular analysis of local coordinating spheres around a given type of central atom.  The \ce{Al} sublattice NNA was performed according to the method in Ref. \cite{Cyster2021} by counting the number of O atoms in the nearest-neighbor shell around the central Al atom for a cutoff radius of 2.4$\:\text{\AA}$ (at $T=900$ K). This specific cutoff radius was chosen to include all neighboring atoms contributing to the first neighbor peak in the \ce{Al}-\ce{O} Pair Distribution function.\

\section{Results and discussion}

\subsection{Thermal Stability and Defect Structure of Bulk $\gamma$-\ce{Al2O3}}

\begin{figure}[H]
    \centering
    \includegraphics[width=\textwidth]{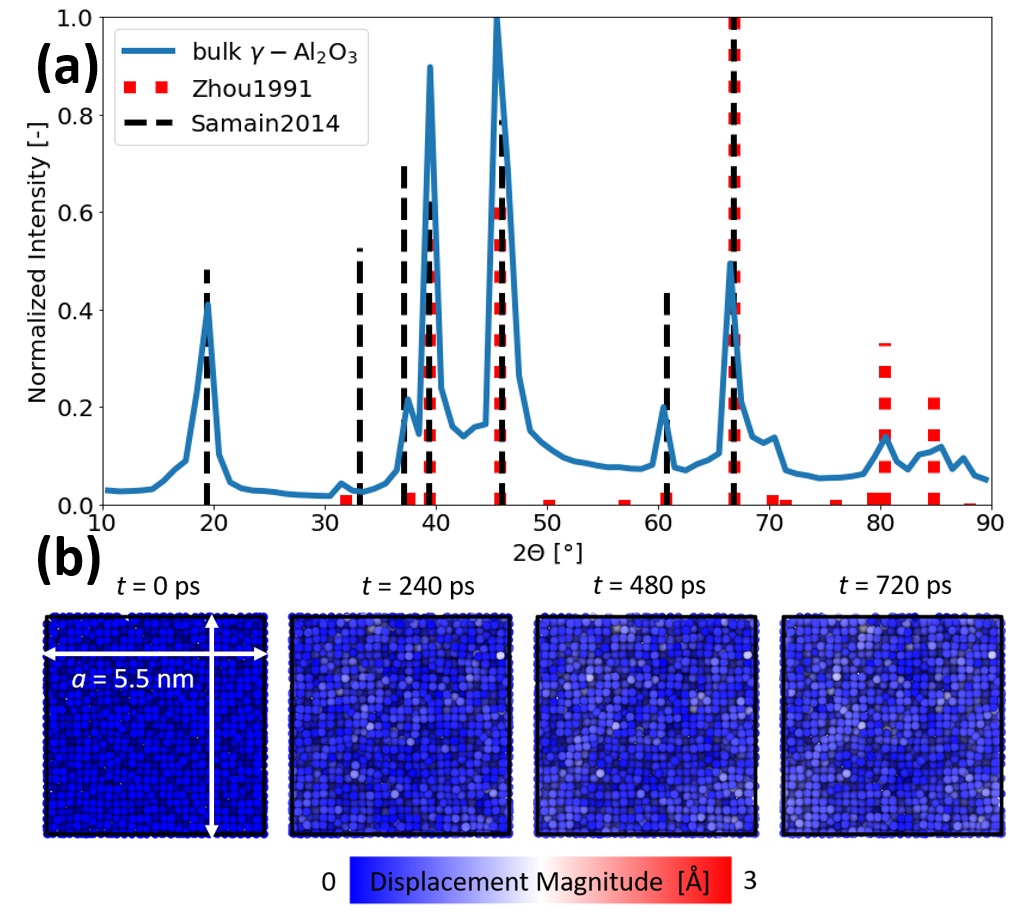}
    \caption{\textbf{(a)} Virtual XRD diffractogram of the simulated $\gamma$-\ce{Al2O3} crystal lattice after isothermal annealing for 720 ps at $T=900$ K (in blue). Corresponding reflections in $\gamma$-\ce{Al2O3}, as reported by the experiment, are indicated by the red \cite{Zhou1991} and black \cite{Samain2014} dashed lines.
    \textbf{(b)} Initial configuration at $t = 0$ ps and maps of the displacement magnitudes inside the $\gamma$-\ce{Al2O3} crystal at $T=900$ K for different simulations times $t = 240, 480, 720$ ps. }
    \label{fig:xrd_disp_rdf}
\end{figure}

Experimental investigations indicate that $\gamma$-\ce{Al2O3} is metastable (in a local energy minimum) below about 1000 K with its defect structure depending on the heating rate and annealing temperature \cite{Wilson1980,Levin1998}. Accordingly, to evaluate the thermal stability and defect structure of bulk $\gamma$-\ce{Al2O3}, the constructed bulk crystal was equilibrated for $t= 720$ ps at $T=900$ K, which is still about 100 K below the bulk transition temperature to $\alpha$-alumina, but close to the experimental bulk transformation temperature of $\gamma$- to $\delta$-\ce{Al2O3} \cite{Zhou1991}. A virtual XRD of the thermally equilibrated $\gamma$-\ce{Al2O3} structure is shown in Fig. \ref{fig:xrd_disp_rdf}\textbf{a}. The peak positions for the simulated structure agree very well with the experimental structure of $\gamma$-\ce{Al2O3}, as reported by XRD \cite{Zhou1991, Samain2014}.\

To verify that a $\gamma$-\ce{Al2O3} structure is metastable at 900 K, the magnitudes of the atomic displacements after 240 ps, 480 ps and 720 ps were visualized by the respective color maps as shown in Fig. \ref{fig:xrd_disp_rdf}\textbf{b}. Indeed, no significant atomic displacements were observed after 720 ps (i.e. the computed atomic displacements remain less than 3$\:\text{\AA}$). Additionally, the FCC oxygen sublattice remains stable during this time interval. Minor atomic displacements are attributed to thermal fluctuations and some diffusivity in the \ce{Al} sublattice due to the presence of vacancies in the tetrahedral interstitial sites. \

\begin{figure}[H]
    \centering
    \includegraphics[width=\textwidth]{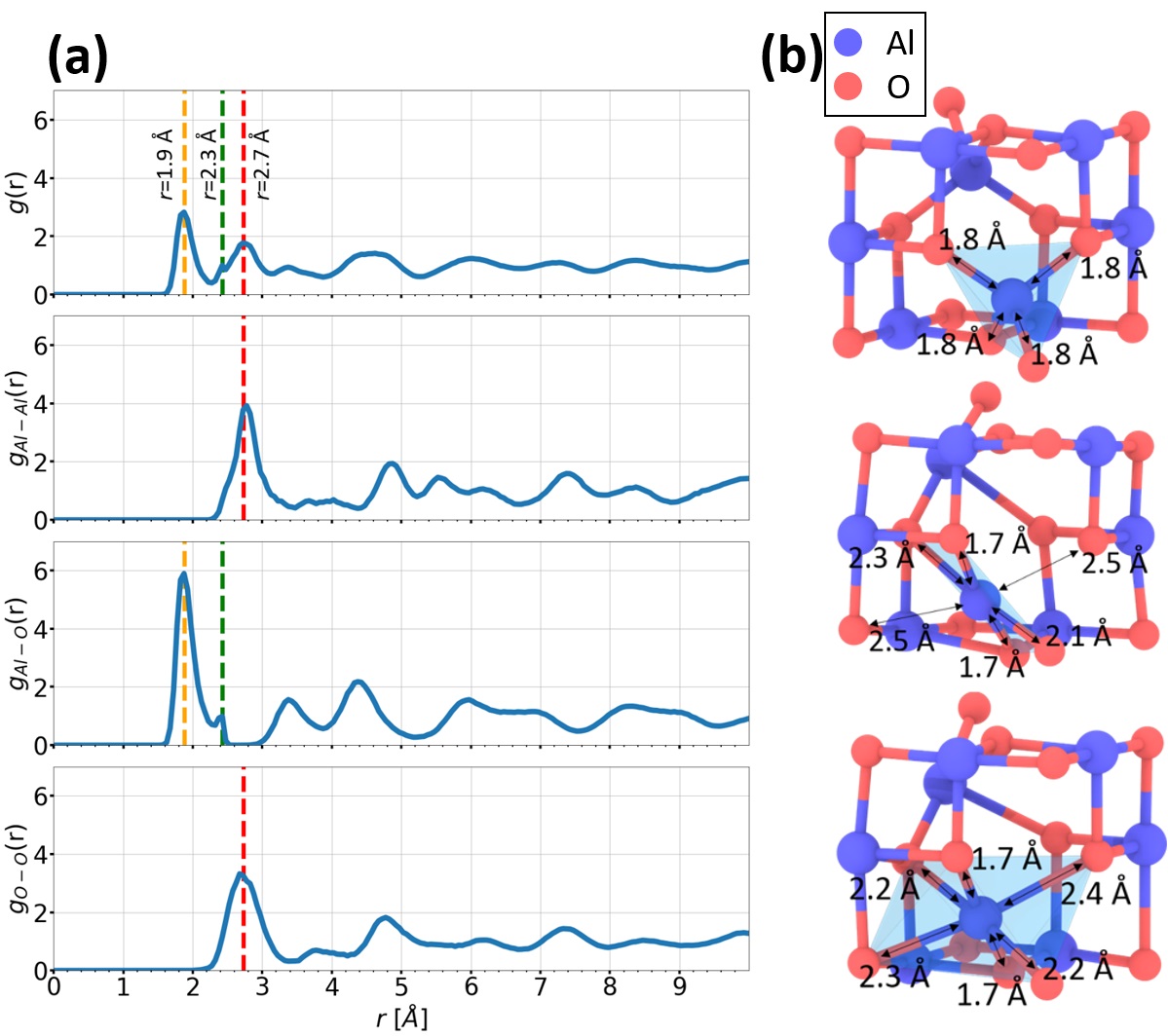}
    \caption{\textbf{(a)} The radial pair distribution function (RDF), $g(r)$, and its respective partial RDFs, i.e. $g_{\text{Al-Al}}(r)$, $g_{\text{Al-O}}(r)$ and $g_{\text{O-O}}(r)$, for bulk $\gamma$-\ce{Al2O3} after thermal equilibration for 720 ps at 900 K. The colored vertical lines indicate the first and second nearest neighbors.
    \textbf{(b)}  Successive positions (from top to bottom) occupied by an \ce{Al} cation originally localized in a tetrahedral site of the $\gamma$-\ce{Al2O3} crystal lattice at 300 K. The intermediate step shows a local distortion of the FCC O sublattice due to the migration of a tetrahedrally coordinated \ce{Al} cation. In the thermally equilibrated structure (at 900 K), the \ce{Al} atom sits in a deformed octahedron with four neighbors at a distance of around 2.2-2.4$\:\text{\AA}$ and two neighbors at 1.7$\:\text{\AA}$. The electrostatic potential energy $U_E$ of the tetrahedral coordinated Al atom slightly increases by 5\% from the top configuration to the bottom configuration. The red atoms are \ce{O} anions; blue atoms are \ce{Al} cations in octahedral and tetrahedral sites.}
    \label{fig:octa_tetra_trans}
\end{figure}

The radial distribution function (RDF) is a very powerful tool for accessing short- and long-range order of the constituent atoms during thermal equilibration of the $\gamma$-\ce{Al2O3} crystal. The extracted RDFs for the simulated $\gamma$-\ce{Al2O3} crystal after thermal equilibration at 900 K are shown in Fig. \ref{fig:octa_tetra_trans}\textbf{a}; the top panel represents the calculated RDF of all constituent atoms, $g(r)$, whereas the lower panels correspond to the partial RDFs for the Al-Al, Al-O, and O-O bonds. The first-order peak(s) for a given chemical bond (i.e. Al-Al, O-O, or Al-O) in the RDF spectrum correspond to the average nearest-neighbor distance(s) (as indicated by colored vertical lines in Fig. \ref{fig:octa_tetra_trans}\textbf{a}), whereas the corresponding higher-order peaks are indicative of long-range order.

As follows from Fig. \ref{fig:octa_tetra_trans}\textbf{a}, the most intense first-order RDF peak at $r\approx 1.9 \:\text{\AA}$ corresponds to the shortest \ce{Al}-\ce{O} bond lengths in $\gamma$-\ce{Al2O3} at 900 K, originating from octahedrally coordinated Al cations ($r\textsubscript{Al-O}\approx 1.94 \:\text{\AA}$) and tetrahedrally ($r\textsubscript{Al-O}\approx 1.80 \:\text{\AA}$) coordinated Al cations in the O-sublattice. The second most distinctive RDF peak at $r\approx 2.7 \:\text{\AA}$ arises from a superposition of the nearest neighbour \ce{Al}-\ce{Al} and \ce{O}-\ce{O} bond distances. These simulated \ce{Al}-\ce{O} and \ce{O}-\ce{O} bond lengths for $\gamma$-\ce{Al2O3} (i.e. $r\textsubscript{Al-O}\approx 1.9 \:\text{\AA}$ and $r\textsubscript{O-O}\approx 2.7 \:\text{\AA}$ at 900 K) agree well with the reported experimental values in the range of $r\textsubscript{Al-O}=1.8-2.2\:\text{\AA}$ and $r\textsubscript{O-O}=2.7-2.9\:\text{\AA}$, respectively \cite{Zhou1991, Lizarraga2011}.

An interesting feature in the $g_{\text{Al-O}}(r)$ spectrum is indicated by the green dashed line at $r\textsubscript{Al-O}\approx 2.3\:\text{\AA}$. As illustrated by the sequence of atomic configurations during thermal equilibration in Fig. \ref{fig:octa_tetra_trans}\textbf{b}, this RDF peak originates from a distortion of the FCC O-sublattice, resulting in the appearance of a characteristic \ce{Al}-\ce{O} bond distance around $r\textsubscript{Al-O}\approx 2.3\:\text{\AA}$. At the beginning of the simulation, the \ce{Al} atoms in the tetrahedral interstices are all positioned in the center of the tetrahedron defined by its four neighboring O atoms (with $r\textsubscript{Al-O}\approx1.8\:\text{\AA}$), as indicated by the light blue shape in the upper configuration of Fig. \ref{fig:octa_tetra_trans}\textbf{b}. During the thermal equilibration procedure, the tetrahedral Al atoms move out of this tetrahedral position in an attempt to increase their coordination with next-neighboring O atoms, as illustrated by the intermediate configuration in Fig. \ref{fig:octa_tetra_trans}\textbf{b}. The final transition state at the end of the thermal equilibration procedure can be visualized by Al in a distorted octahedral environment (further referred to as "quasi-octahedron") with four O nearest-neighbors at $r\textsubscript{Al-O}\approx2.2-2.4\: \text{\AA}$ and two O neighbors at $r\textsubscript{Al-O}\approx1.7\: \text{\AA}$, as indicated by the light blue shape in the bottom configuration of Fig. \ref{fig:octa_tetra_trans}\textbf{b}. The formation of these quasi-octahedral Al coordination spheres results in the appearance of an additional nearest-neighbor peak centered at 2.3$\:\text{\AA}$ in the corresponding RDFs of the bulk $\gamma$-alumina crystal in Fig. \ref{fig:octa_tetra_trans}\textbf{a}. The simulations at 900 K thus evidence a driving force for tetrahedral Al cations to increase their coordination by O neighboring atoms, which is a direct consequence of the metastable nature of bulk $\gamma$-\ce{Al2O3} at elevated temperatures of 900 K, as discussed in the following.

Local distortions of the O-sublattice in $\gamma$-\ce{Al2O3} due to small displacements of tetrahedral Al cations are very common for the transformation of $\gamma$-\ce{Al2O3} to $\alpha$-\ce{Al2O3} (with $\delta$-\ce{Al2O3} and $\theta$-\ce{Al2O3} as intermediate transition phases) \cite{Zhou1991}. The driving force for such lattice distortions is the tendency to increase the coordination of tetrahedral Al cations by O anions \cite{Blonski1993}. The \emph{local coordination sphere} (LCS) of a given atom in a solid describes its short-range order and comprises the number and type, as well as angles and distances, between the central atom and its neighboring atoms. In this regard, numerous studies have reported "distorted" local coordination spheres of Al (further referred to as Al-LCS) that differ from the well-defined 4-fold and 6-fold Al-LCS in crystalline and amorphous alumina polymorphs. \cite{Meinhold1993,Kunath-Fandrei1995,MacKenzie1999,Lee2014,Shi2019,Xu2021} However, such distorted Al-LCS have been very controversially discussed in the literature up to date. \cite{Blonski1993, MacKenzie1999,Wang1999,Lizarraga2011,Zhou2019} Most \textsuperscript{27}Al nuclear magnetic resonance (NMR) studies of \textit{am}-\ce{Al2O3} and $\gamma$-\ce{Al2O3} have usually attributed intermediate coordination spheres between 4 and 6 (as associated with a resonance in the range of 30-35 ppm) to pentacoordinated Al, but without specifying characteristic bond lengths and angles between Al and its neighboring O anions (as provided for the quasi-octahedron in Fig. \ref{fig:octa_tetra_trans}\textbf{b}).\cite{Meinhold1993,Kunath-Fandrei1995,MacKenzie1999,Lee2014,Shi2019,Xu2021} Experimental studies by XRD \cite{Zhou2019, MacKenzie1999}, as well as theoretical studies by MD \cite{Blonski1993,Wang1999,Lizarraga2011}, rather evidence the existence of quasi-octahedral Al-LCS, similar to the one observed in the present study (see Fig. \ref{fig:octa_tetra_trans}\textbf{b}). Alternative geometries, such as elongated bonds in a distorted AlO\textsubscript{4} tetrahedron, as well as an Al cation in an octahedral interstitial site of the O FCC sublattice with an adjacent O vacancy, have also been proposed to give rise to the characteristic "5-fold" NMR resonance \cite{Meinhold1993, MacKenzie1999}. This poses the question if the quasi-octahedral Al-LCS found in our simulations could give rise to a similar \textsuperscript{27}Al NMR resonance. Such verification is out of the scope of this study but would be an interesting hypothesis to be cross-checked by NMR theory. Notably, 5-fold (i.e. quasi-octahedral \ce{Al} spheres) are believed to be key structural building blocks of \textit{am}-\ce{Al2O3} \cite{Davis2011,Shi2019}, as is also evidenced in our simulations of structural disorder in $\gamma$-\ce{Al2O3} NPs upon annealing.\

To summarize, our model predictions of quasi-octahedral Al-LCS in bulk $\gamma$-\ce{Al2O3} at 900 K are in excellent agreement with the wealth of experimental observations showing that a bulk transformation of $\gamma$-\ce{Al2O3} to $\alpha$-\ce{Al2O3} (i.e. a transformation of the O sublattice from FCC to HCP with all Al cations positioned in octahedral interstices) is kinetically hindered at 900 K, although local distortions of the O sublattice by displacement of tetrahedral Al cations are thermally activated and thus proceed with increasing temperature and time.\

To further investigate the origin of the formation of distorted octahedrons in the present study, the electrostatic potential energies of a central Al cation in the tetrahedral and quasi-octahedral Al-LCS are a function of the local atomic environment and were calculated from
\begin{align}
    U_E(r)=k_{\text{e}}q_{\ce{Al}}\sum_{i=1}^{n_{\ce{O}}}\frac{Q_i}{r_i}
\end{align}
with $k_{\text{e}}$ being Coulombs constant; $q_{\ce{Al}}$ and $Q_i$ being the charges of the central Al cation and its neighboring \ce{O} atom; $r_i$ is the distance between the \ce{Al} atom and \ce{O} atom $i$. The thus calculated potential energy is higher for the quasi-octahedral Al-LCS, which suggests that the lattice distortion is not driven by ionic bonding and Coulombic forces and thus must have a different physical-chemical origin. Interestingly, the Al-O bond angles of the central \ce{Al} change from the typical tetrahedral value of 109.5\textdegree{} to a value around 90\textdegree{}, characteristic of octahedral coordination. This indicates that the transformation from tetrahedral to distorted octahedral coordination might be driven by the covalent bonding characteristics of \ce{Al2O3} \cite{Sun2008}. The covalent bond is known to have a directional character: i.e. the preferred bond angles of the central atoms are dependent on the orientation of its neighboring atoms that share an electron pair. Due to the electronegativity difference between \ce{Al} and \ce{O}, the \textit{van Arkel-Ketelaar Triangle} predicts a mixture of covalent and ionic bonding characteristics for alumina \cite{Allen1993}. Indeed, DFT calculations have shown that the \ce{Al}-\ce{O} bond has a weak, but distinct covalent contribution, whereas the respective \ce{O}-\ce{O} bond has no covalent bond character \cite{Rahane2011,Menendez-Proupin2005}. Such partial covalent bonding characteristics were shown for \textit{am}-, $\gamma$- and $\alpha$-\ce{Al2O3} \cite{Menendez-Proupin2005}. With increasing temperature, the influence of the covalent bond characteristic of \ce{Al2O3} becomes more dominant \cite{French1994}, thus activating the formation of quasi-octahedral Al-LCS. The partial covalency of the Al-O bond, on the one hand, favors the formation of defined 4-fold Al[O\textsubscript{4}] and/or 6-fold Al[O\textsubscript{6}] corner-sharing polyhedral building blocks, resulting in short-range order. The lack of covalency of the O-O bond, on the other hand, allows a "flexible" modification of the interconnected network of these corner-sharing polyhedral building blocks by local distortions of the O-sublattice \cite{Zachariasen1932, Snijders2005, Shi2019} (thus affecting the long-range order, i.e. the translational symmetry and periodicity of the atomic arrangements). Accordingly, this phenomenon in amorphous oxides is often referred to as bond flexibility \cite{Zachariasen1932, Revesz1981, Jeurgens2000, Snijders2005}.

Our simulations are in excellent agreement with experiment (see above), but contradict certain \textit{ab initio} studies which predict that \ce{Al} is energetically more favorable in the tetrahedral interstices of the FCC O-sublattice instead of the octahedral ones (thus suggesting that the respective cation \emph{vacancies} preferably occupy the octahedral interstices \cite{Wolverton2001,Gutierrez2002a,Pinto2004}). However, such \textit{ab initio} studies generally operate on extremely small supercells with tens or maximum hundreds of atoms, which does not allow to access variation in long-range ordering phenomena and may also introduce size effects associated with intrinsic vacancies in the Al sublattice of $\gamma$-\ce{Al2O3}. Even more importantly, first-principle calculations (performed at zero Kelvin) do not account for thermal effects, which play a key role in activating local lattice distortions over longer distances within relatively short simulation times. The COMB3 potential can be operated on much larger simulation cells with thousands of atoms, which provides access to such long-range structural relaxations at finite temperatures, which is crucial for tracing the thermal stability of alumina transition polymorphs: i.e. to allow predictions of key structural building blocks that are commonly observed experimentally, such as the quasi-octahedral Al-LCS. The distribution of intrinsic cation vacancies and the induced local lattice distortions are associated with long-range strain-fields. These elastic fields play an important role in shaping the structural properties of $\gamma$-Al$_2$O$_3$ and are important for a deeper understanding of the material's properties and performance. At the same time, it is possible to neglect long-range Coulomb interactions, as they are mostly canceled out by the shielding effects, often allowing for the description of partially or fully ionic solid materials with short-range interatomic potentials. For example, recent studies have shown that neural network potentials can provide extremely accurate predictions for glasses such as Al$_4$P$_2$O$_7$, only by fitting local environments, without considering atomic charges and electrostatic interactions\cite{Batzner2022new,Li2020,Vandermause2020new}. However, as we will demonstrate later in this paper, the long-range electrostatic interactions become essential in describing long-range structural relaxations correctly in systems with free surfaces such as nanoparticles.\

\subsection{Phase Stability of $\gamma$-\ce{Al2O3} Nanoparticles}

\begin{figure}[H]
    \centering
    \includegraphics[width=\textwidth]{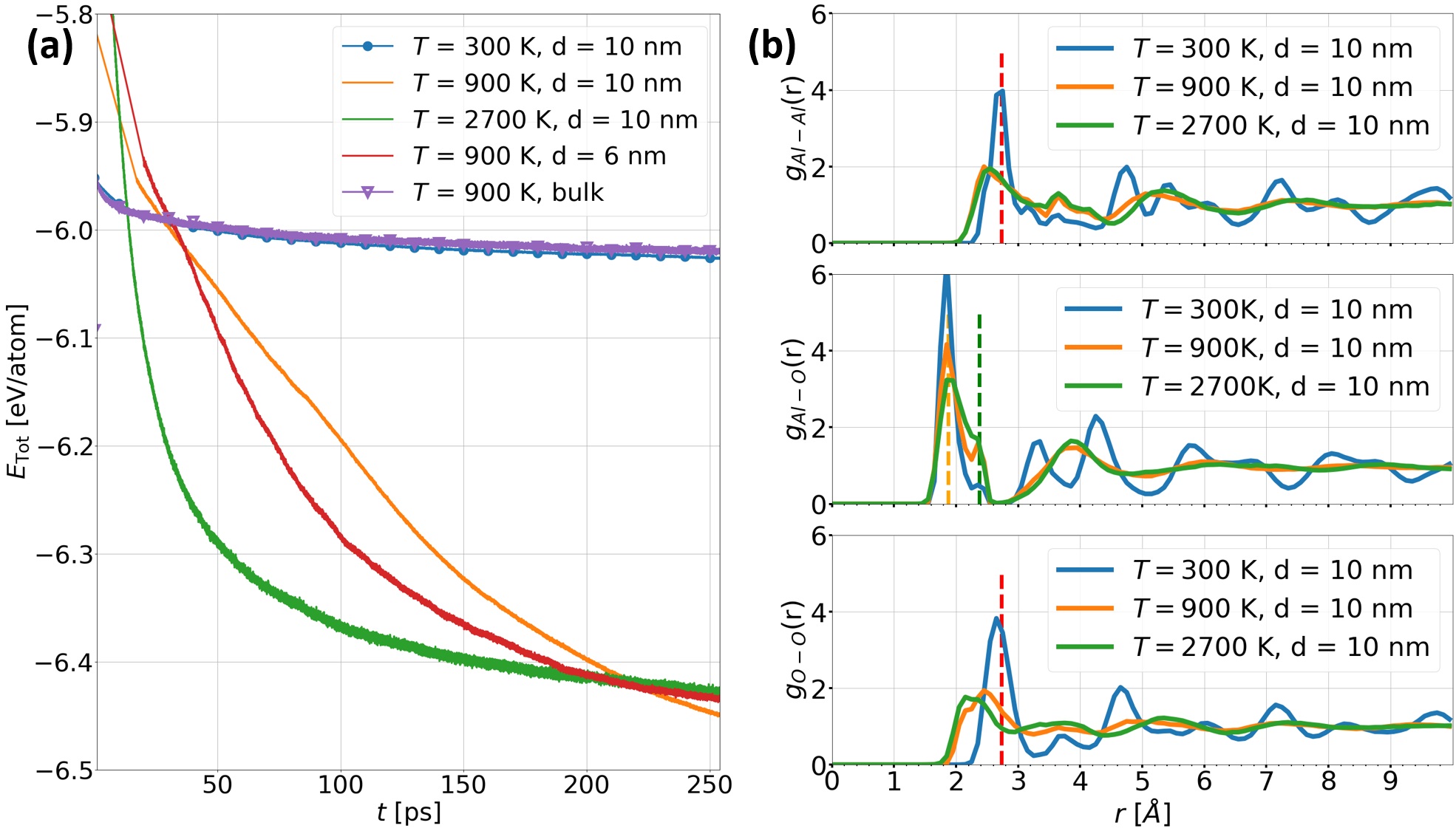}
    \caption{\textbf{(a)} The potential energy $E_{\text{tot}}$/atom of a $d=$ 6,10 nm NP and bulk crystal of $\gamma$-\ce{Al2O3} during thermal equilibration for 250 ps at a constant temperature of \emph{T} = 300, 900 and 2700 K.
    \textbf{(b)} The NP structure after thermal equilibration is visualized by the corresponding Al-Al, Al-O, O-O partial radial distribution functions, $g(r)$, of the NP in its final state (i.e. after 250 ps of thermal equilibration). Note that the RDFs were extracted from the core of the NPs, 1 nm away from the free surface. For an indication of the peaks, it is referred to the RDFs of bulk $\gamma$-\ce{Al2O3} in Fig. \ref{fig:octa_tetra_trans}\textbf{a}.}
    \label{fig:equilibration}
\end{figure}

The calculated thermal stability and defect structure of bulk $\gamma$-\ce{Al2O3} at 900 K can be compared with the predicted (size-dependent) phase stability of $\gamma$-\ce{Al2O3} NPs at 300 K, 900 K and 2700 K. Two complementary observables were used to study the metastable phase stability of a 10 nm spherical $\gamma$-\ce{Al2O3} NP during thermal equilibration at 300 K, 900 K and 2700 K (i.e. above the bulk melting point of $\gamma$-\ce{Al2O3} at about 2328 K). First, to investigate whether a metastable NP state is reached at the end of the simulation, the evolution of the potential energy per atom, $E_{\text{Tot}}$, of the respective NP (or bulk crystal) was monitored as a function of the simulation time: see Fig. \ref{fig:equilibration}\textbf{a}. Next, the NP structure reached at the end of the thermal equilibration (i.e. after 250 ps) was compared to that of the bulk crystal by analyzing the corresponding partial RDFs, $g(r)$, in the final state: see Fig. \ref{fig:equilibration}\textbf{b} (to be compared to the RDF of the bulk crystal in Fig. \ref{fig:octa_tetra_trans}\textbf{a}). To reveal the nanosize effect on the metastable phase stability of the $\gamma$-\ce{Al2O3} NPs, the simulations at $T = 900$ K were performed for two different NP diameters of $d = 6$ nm and $d = 10$ nm.\

The following main conclusions can be drawn from Fig. \ref{fig:equilibration}:

\begin{enumerate}[\itshape(i)]

\item The potential energy evolution of the bulk crystal at $T = 900$ K is equivalent to that of the 10 nm NP at $T = 300$ K, both reaching a metastable state within a few picoseconds.

\item At $T = 900$ K, the average potential energies of the 6 nm and 10 nm NPs initially decrease very rapidly with increasing time, indicating a structural transformation inside the NPs during thermal equilibration. The slope of the potential energy curve gradually flattens and eventually reaches a slope, similar to the metastable states of the bulk crystal at 900 K, the solid NP at 300 K, and the liquid NP at 2700 K. This indicates that the phase transformation inside the NP is largely completed towards the end of the simulation (i.e. a metastable state is reached).

\item For the highest temperature of $T = 2700$ K (i.e. above the bulk melting point of alumina), the total energy of the 10 nm NP decreases rapidly during the initial stage of equilibration and then becomes approximately constant after 150-200 ps. This suggests that the NP is molten and a thermodynamically-preferred equilibrium liquid state is reached within the simulation time period.

\end{enumerate}

\textbf{Ad(\textit{i}):} The RDFs for the metastable states of the core of the 10 nm NP (i.e. disregarding 1 nm NP shell for the RDF analysis) at 300 K and the bulk crystal at 900 K are indeed very similar: compare Figs. \ref{fig:octa_tetra_trans}\textbf{a} and \ref{fig:equilibration}\textbf{b}. The corresponding nearest-neighbor peaks of $g_{\text{Al-O}}(r)$ and $g_{\text{O-O}}(r)$ are located at the same position and have similar peak maxima. Only a slight difference of less than $0.1\:\text{\AA}\:$ in the position of the second nearest-neighbor shell in $g_(r)$ is noticed, originating from the thermal expansion of the $\gamma$-\ce{Al2O3} crystal at 900 K. The RDF peak at $r\textsubscript{Al-O}\approx 2.3\:\text{\AA}$, as representative for quasi-octahedral (or "5-fold") Al-LCS, is slightly more pronounced for the 10 nm NP at 300 K than for the bulk crystal at 900 K (i.e. even though the NP experiences a much lower temperature). This suggests that the distortion of the long-range periodicity FCC O-sublattice by the formation of quasi-octahedral Al-LCs can proceed at RT for the $\gamma$-\ce{Al2O3} NP, whereas such lattice distortions require a much higher temperature for the respective bulk crystal. The free surface of the NP aids to overcome the energy barrier for these lattice distortions at temperatures as low as RT. In the absence of a free surface (i.e. for the bulk crystal), such lattice distortions are kinetically hindered at 300 K. This is consistent with experimental findings for transition alumina phases, showing that deviating Al-LCS environments (i.e. different from the well-known 4-fold or 6-fold Al-LCS) preferentially arise in the vicinity of the surface, acting as key catalytically active surface sites for heterogeneous catalysis.\cite{Blonski1993,Lee2014,Khivantsev2021} \

\textbf{Ad(\textit{ii}):} The structural transformation of 6 nm and 10 nm NPs at 900 K can be divided into two different kinetic stages (based on the evolution of the potential energy over time). For example, for the 10 nm NP, a rapid decrease from the initial value to -6.3 eV/atom in about 150-200 ps is followed by a second stage with a slower decrease of the potential energy per atom, approaching a slope similar to those of the metastable states of the bulk crystal at 900 K and the 10 nm NP at 300 K, as well as of the equilibrium state of the molten NP at 2700 K.
As follows from Fig. \ref{fig:equilibration}\textbf{a}, the initial rate of structural transformation is faster for the smaller NP due to its higher surface-to-volume ratio. Consequently, the smaller NP reaches a lower energetic state with slow kinetics (i.e. a flatter slope of the potential energy with time) within shorter equilibration times than the larger NP. Its $g(r)$ - not shown here - shows a lack of long-range order, similar to those of the molten NP at $T=2700$ K (see below), thus indicating a pronounced amorphization of the surface region. Furthermore, the quasi-octahedral ("5-fold") peak  intensity of the 10 nm NP (at $r\textsubscript{Al-O}\approx 2.3\:\text{\AA}$) increases with increasing temperature from 300 K to 2700 K, as accompanied by a proportional decrease of the overlapping 4-fold and 6-fold peak intensities (at $r\textsubscript{Al-O}\approx 1.9\:\text{\AA}$). This evidences that the formation of quasi-octahedral Al-LCS is thermally activated (i.e. associated with an energy barrier), in accordance with the above findings for the $\gamma$-\ce{Al2O3} crystal.

\textbf{Ad(\textit{iii}):} The partial RDFs of the NPs at 900 K and 2700 K (see Fig. \ref{fig:equilibration} \textbf{b}) show the persistence of the first nearest-neighbor peak(s), but strong damping or nearly vanishing of the higher-order peaks. This indicates a lack of long-range order (i.e. a loss of translational symmetry and periodicity of the atomic arrangements), which is characteristic of liquid and amorphous oxide phases (which can be treated as configurationally frozen liquids \cite{Jeurgens2000}). Indeed, the nearest-neighbor peak at $r\textsubscript{Al-O}\approx 1.9 \:\text{\AA}$ matches well with the respective peak in the partial RDF of molten alumina at $T = 2500$ K, as determined by diffraction \cite{Landron2001a, Shi2019}, as well as with the simulated RDFs of liquid alumina (by \textit{ab initio} MD) \cite{Jahn2007}.\

In particular, the quasi-octahedral (or 5-fold) Al-LCS peak at $r\textsubscript{Al-O}\approx2.3$ \AA$\,$ increases with increasing temperature (as well as, to a lesser extent, with decreasing NP size; see above) and is very pronounced for the amorphous NP at 900 K and the liquid NP at 2700 K. This implies that quasi-octahedral (or "5-fold") Al-LCS is an important building block of amorphous and liquid alumina, in agreement with experiment \cite{Kunath-Fandrei1995,Landron2001a, Shi2019} and theory \cite{Blonski1993,Lizarraga2011, Shi2019}. The increase in the intensity of the quasi-octahedral Al-LCS peak with increasing temperature (and decreasing NP size) is accompanied by a broadening of the distributions of the O-O and Al-Al nearest-neighbour bond lengths (see Fig. \ref{fig:equilibration}\textbf{b}). In particular, the nearest-neighbor peak in $g_{\text{O-O}}(r)$ of the 900 K and 2700 K equilibrated NPs develops a pronounced shoulder towards lower bond lengths. Substantial broadening of the O-O nearest-neighbor bond length distribution in combination with the thermally activated formation of quasi-octahedral Al-LCS evidences a progressive distortion of the FCC O-sublattice by the insertion of "flexible" (i.e., network-modifying \cite{Zachariasen1932}) "Al[O\textsubscript{5}]" building blocks into the initial network of interconnected Al[O\textsubscript{6}] and Al[O\textsubscript{4}] polyhedra.\

In summary, our simulations indicate that $\gamma$-\ce{Al2O3} NPs with diameters of 6 nm and 10 nm are metastable at 300 K. A crystalline-to-amorphous transformation characterized by loss of long-range order sets in toward elevated temperatures. This $\gamma\rightarrow{am}$-\ce{Al2O3} transition reaches a metastable state in less than 200 ps at 900 K. To our knowledge, no other theoretical studies have investigated the metastable phase stability of alumina NPs with sizes as large as $d = 10$ nm in the range from 300 K up to 2700 K. Laurens \textit{et al.} compared NPs with $d < 12$ nm only at $T = 300$ K; their model calculations predict a thin, size-independent amorphous oxide shell around $\gamma$-\ce{Al2O3} NPs at RT, but the evolution of the core-shell structure toward elevated temperatures was not studied.\cite{Laurens2020}\

\subsection{Amorphization of the NP}
\label{subsec: NP amorphization}

\begin{figure}
    \centering
    \includegraphics[width=\textwidth]{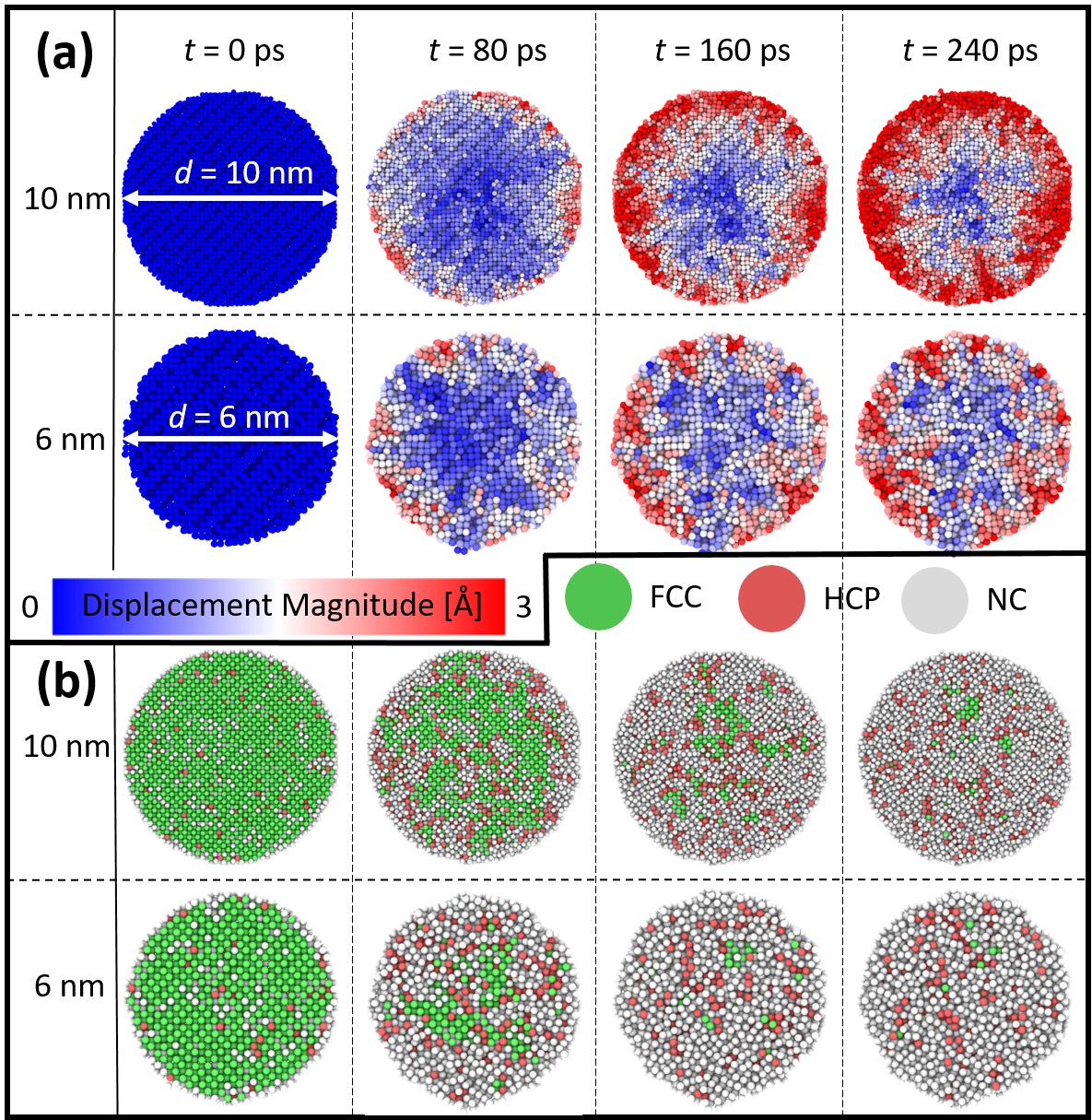}
    \caption{
    \textbf{(a)}Displacement magnitudes in the cross-section of a $d=10$ nm NP (top) and $d=6$ nm NP (bottom) at $T=900$ K. The reference positions for the atomic displacements were taken at $t=0$.
    \textbf{(b)} Snapshots in the same cross-section of the oxygen sublattice in the NPs at different times. Oxygen atoms are colored according to the \textit{Ackland and Jones} analysis.
    }
    \label{fig:disp_AJA}
\end{figure}

As discussed in the previous section, the $\gamma$-\ce{Al2O3} NPs experience a loss of long-range order at 900 K, characteristic of the development of an amorphous oxide phase. To investigate the NP amorphization process during thermal equilibration at 900 K in more detail, the corresponding atomic displacements in the cross-section over simulation intervals of 80, 160, and 240 ps were studied (see Fig. \ref{fig:disp_AJA}\textbf{a}). Therefore, the atomic displacements (with reference to \textit{t} = 0) are highest at the surfaces of the NP, in accordance with the preferred formation of quasi-octahedral Al-LCS at the NP surface. The structural relaxations propagate from the surface into the core of the NPs, in a 'columnar' manner. The visibility of propagation mainly depends on the crystallographic direction. This effect is a result of the anisotropic nature of surface stress and surface segregation which are discussed in the last section.\

The transformation of the FCC O-sublattice of $\gamma$-\ce{Al2O3} during amorphization was tracked using AJA: see Fig. \ref{fig:disp_AJA}\textbf{b}. Both NPs have the characteristic FCC O-sublattice of undistorted $\gamma$-\ce{Al2O3} at the beginning of the simulation. At $T=900$ K, the number of FCC atoms decreases during thermal equilibration, indicative of local distortions of the O-sublattice due to the formation of quasi-octahedral Al-LCS. After 250 ps, the vast majority of surface O anions cannot be identified as FCC and are thus labeled as non-crystalline (NC). The O lattice distortions nucleate at the surface, since the surface is already partially reconstructed during the equilibration of the initial state at 300 K in order to lower the energy of the unrelaxed polar oxide surface.\cite{Jeurgens2000, Reichel2008b} As a result, the activation barrier for the $\gamma\rightarrow{am}$-\ce{Al2O3} transformation is lowest at the NP surface.\

According to bulk thermodynamics, $\alpha$-Al$_2$O$_3$ is the only stable bulk phase in the Al-O system \cite{Wefers1987}. The transformation from $\gamma$- to $\alpha$-alumina requires the transformation of the oxygen sublattice from an FCC to an HCP structure, which is not a local transformation and requires overcoming a high energy barrier. Indeed, experimental works show that all bulk alumina transition phases eventually transform into $\alpha$-Al$_2$O$_3$ at T > 1400 K, independently of their synthesis pathway. Therefore, in the current study, all other polymorphic phases of Al$_2$O$_3$ are designated as \textit{metastable}. Depending on the temperature, different thermal activation barriers may be surmounted with increasing simulation time to reach different local minima (corresponding to different transition alumina phases) or eventually the global minimum (corresponding to $\alpha$-Al$_2$O$_3$, only reachable at T > 1400 K). This implies that our simulations of $\gamma$-Al$_2$O$_3$ crystals and NPs are exploring configurations around local minima (in the direction of the global minimum) during the simulation which can be reached or not as dictated by temperature. However, as discussed in the Introduction, for NPs with $d<12$ nm, $\gamma$--\ce{Al2O3} and/or \textit{am}-\ce{Al2O3} become the thermodynamically preferred alumina phases. Indeed, in our MD simulations, O anions only experience an HCP environment on their transformation to NC. After 250 ps of thermal equilibration, both NPs end up with all O atoms being classified as NC, indicating a complete loss of long-range order in the FCC O-sublattice with the remaining short-range order dictated by the randomly interconnected Al[O]$_{n}$-polyhedra.\

\begin{figure}
    \centering
    \includegraphics[width=\textwidth]{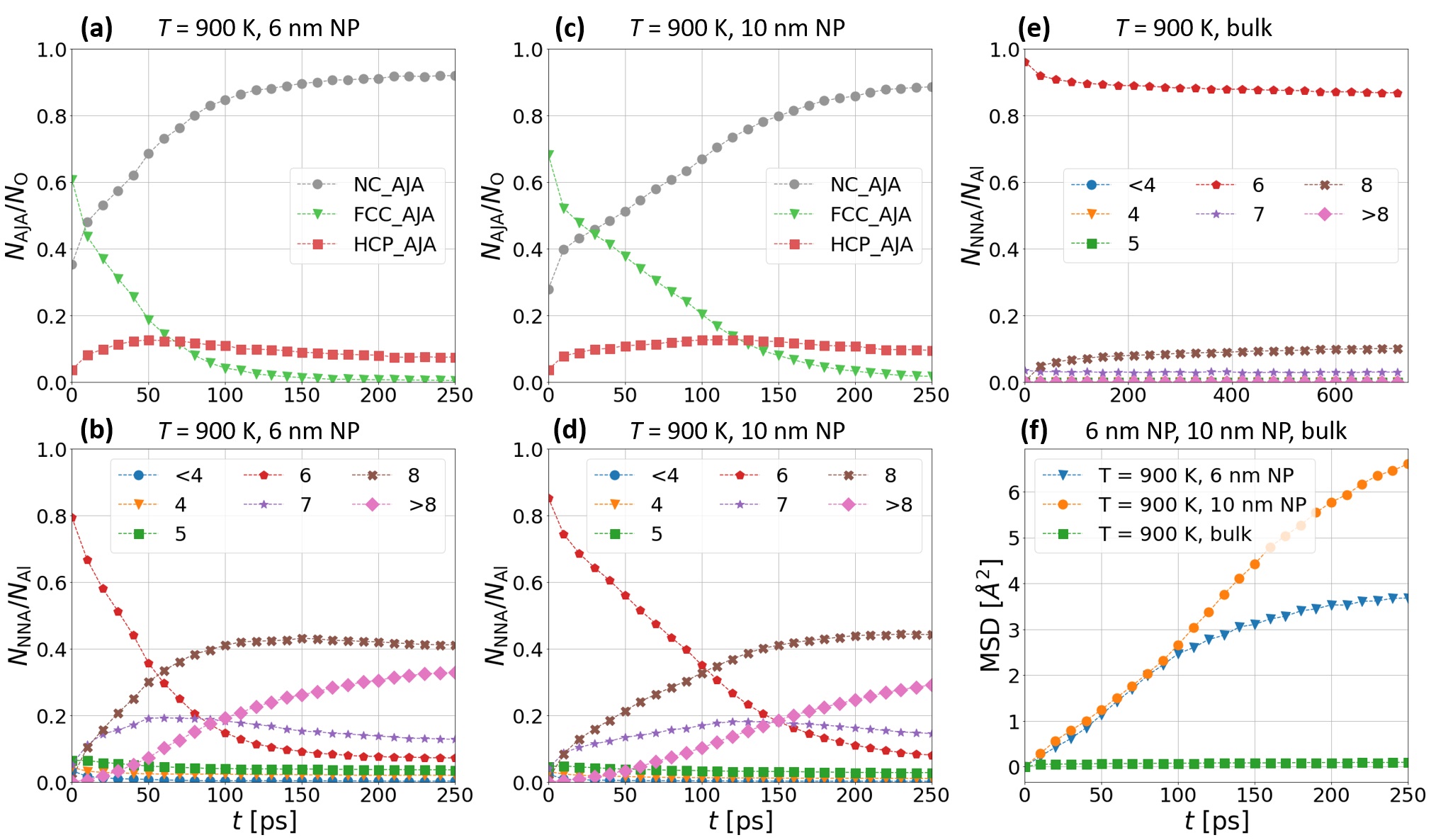}
    \caption{
    \textbf{(a,c)} Ratio of the number of \ce{O}-atoms in the local configuration assigned by the AJA procedure, $N_\text{AJA}$, and the total number of \ce{O}-atoms, $N_\text{O}$, plotted versus simulation time for \textbf{(a)} the $d = 6$ nm NP at $T = 900$ K and \textbf{(c)} the $d = 10$ nm NP at $T = 900$ K.
    \textbf{(b,d,e)} Ratio between the number of \ce{Al}-atoms in the different local coordination spheres evaluated by the NNA procedure, $N_\text{NNA}$, divided by the total number of \ce{Al}-atoms, $N_\text{Al}$, plotted versus simulation time for \textbf{(b)} the $d = 6$ nm NP at $T = 900$ K, \textbf{(d)} the $d = 10$ nm NP at $T = 900$ K and \textbf{(d)} the bulk $\gamma$--\ce{Al2O3} crystal at $T=900$ K.
    \textbf{(f)} The mean-squared displacements of all atom types at $T = 900$ K for the bulk $\gamma$--\ce{Al2O3} crystal, as well as the MPs with $d = 6$ nm and $d = 10$.
    \label{fig:rates}}
\end{figure}

The transformation rates of the NP with $d = 6$ nm and $d = 10$ nm can be compared by tracing the AJA analysis with increasing equilibration time (at 900 K): see Figs. \ref{fig:rates}\textbf{a} and \ref{fig:rates}\textbf{c}, respectively. It follows that the transformation rate from FCC to NC O-atoms is faster and completed in a shorter time for the smaller NP, in accordance with our results and Refs. \cite{Mavric2019,Tavakoli2013,Laurens2020}.\

Application of the NNA procedure allows us to trace the evolution of the different Al-LCS types during thermal equilibration at $T = 900$ K: see Fig. \ref{fig:rates}\textbf{b,d}. It follows that, in the initial state (i.e. after equilibration at 300 K), the vast majority (around 80 \%) of the Al atoms are in octahedral coordination with six O nearest neighbors. All other Al-LCS types (i.e. $<$4, 4, 5, 7, 8, $>$8) initially have a total fraction of less than 0.1 (with 5-fold and 4-fold being the second and third most frequent Al-LCS), very similar to the bulk $\gamma$--\ce{Al2O3} crystal at $T=900$ K (see Fig. \ref{fig:rates}\textbf{e}). The second most frequent 5-fold Al-LCS results from stochiometric cation vacancies. All other Al-LCS in the initial state are attributed to thermal fluctuations, which arise as a result of the defined finite cutoff in the NNA analysis. The fraction of 6-fold Al-LCS in bulk $\gamma$-\ce{Al2O3} remains at a constant level of 0.9 during equilibration at $T = 900$ K (up to $t=$ 720 ps), in accordance with our previous findings that Al cations in bulk $\gamma$-\ce{Al2O3} at $T = 900$ K are either in octahedral or distorted octahedral coordination.

During the $\gamma\rightarrow{am}$-\ce{Al2O3} transformation of the NPs, the number of 6-fold coordinated Al atoms is steadily decreasing, as accompanied by an increase of 7-fold Al-LCS, as well as some higher 8-fold and $>$8-fold Al-LCS (each with fractions up to $\approx 0.3$). All other Al-LCS-types (i.e. $<$4, 4, 5) remain practically constant (adding up to a maximum of $7\%$ of the total number of Al atoms). This indicates that in addition to the quasi-octahedral Al-LCS (as for bulk $\gamma$-\ce{Al2O3} at 900 K), several additional local coordination spheres are formed during the $\gamma\rightarrow{am}$-\ce{Al2O3} transition. In this regard, it is emphasized that vacancies on the cation and anion sublattices are not stable in amorphous oxides; i.e. the creation of a vacancy in an amorphous oxide does not result in a stable unoccupied (sub)lattice site, but will assimilated by local structural relaxations (as aided by the free volume and high bond flexibility of the amorphous structure),\cite{Guo2018,Broqvist2007} as accompanied by the appearance of higher-order Al-LCS types. All Al-LCS-types in the \textit{am}-\ce{Al2O3} NPs that differ from 4- and 6-fold coordination hint at local heterogeneities in the stoichiometric composition, as associated with the creation of local space charge regions, as discussed in more detail in the next section.

The distribution of different Al-LCS-types for the 6 nm NP is reaching a metastable state after 200 ps, which is at least 5 ps slower than the time required to reach a metastable state of the corresponding O-sublattice (see above). This suggests that the local displacements of Al cations are assisted by O-sublattice distortions. A similar trend is observed for the larger NP, although a true metastable state is not reached within the time frame of our simulation. This finding is remarkable since, generally, the transformation from $\gamma$-alumina to $\theta$-alumina is thought to be induced by the movement of interstitial Al cations \cite{Cai2003}. This suggests that a phase transformation between crystalline alumina polymorphs of the same O sublattice type may proceed by a redistribution of interstitial Al cations, whereas a respective $\gamma\rightarrow{am}$-\ce{Al2O3} polymorphic transformation is realized by local structural relaxations through collective displacements of anions and cations.\

Finally, the mean squared displacements were calculated during thermal equilibration as shown in Fig. \ref{fig:rates}\textbf{e}. The respective diffusion coefficients were not calculated since the applied Langevin thermostat is known to dampen the kinetics of a system, affecting the resulting kinetic properties.\cite{Basconi2013a} The mean squared displacements for bulk $\gamma$-\ce{Al2O3} at 900 K can be neglected (see Fig. \ref{fig:xrd_disp_rdf}\textbf{b}), indicating that a bulk metastable state is reached. In contrast, for the NPs, the mean squared displacements increase with time (at 900 K), indicating that the $\gamma\rightarrow{am}$-\ce{Al2O3} phase transformation associated with collective atomic displacements (see above) and surface diffusion is thermally activated. The mean squared displacement for the 6 nm NP becomes much slower after about 150 ps, indicating that a metastable state is reached and the following displacements can be attributed to self-diffusion. For the 10 nm NP, the mean squared displacement still increases substantially with time, indicating that the $\gamma\rightarrow{am}$-\ce{Al2O3} phase transformation has not yet been fully completed within the time frame of our simulation. Interestingly, the mean squared displacement initially increases with the same rate for both NPs, which implies that the velocity of the inwardly moving amorphization front (i.e. from the surface to the core) is independent of the NP size and thus dictated by the bond flexibility of the \textit{am}-\ce{Al2O3} phase. The larger mean square displacement for the larger NP after 250 ps of thermal equilibration is a direct consequence of its larger surface area. Our simulations of the crystalline-to-amorphous transformation of Al$_2$O$_3$ NPs evidence a delicate interplay between lattice distortions, stresses, and space charges. These basic findings may contribute to the advancement of a broad range of real-world applications. For example, in the field of energy storage, amorphous Al$_2$O$_3$ NPs have been used as a component in high-performance lithium-ion batteries due to their high lithium-ion storage capacity\cite{Sheem2012new}. A crystalline-to-amorphous transformation can lead to significant changes in the ionic conductivity and charge storage properties, thus affecting the battery performance. In the field of optics and optoelectronics, Al$_2$O$_3$ NPs have been used as a component in photovoltaic devices such as solar cells\cite{Woo2021new}; a crystalline-to-amorphous transformation will change the electronic properties, thus modifying the energy-conversion efficiency. Oxide NPs are also key ingredients for the biosensing of molecules; the interaction with biomolecules crucially depends on the size, morphology, and (defect) structure of the oxide NPs, as affected by any phase transition\cite{Liu2019new}.\

\subsection{Stress Development in NPs}

It can be assumed that the amorphization of the NPs at 900 K will be affected by the acting surface stresses on the NP.\cite{Holec2021, Pizzagalli2021} Notably, surface stress is typically evaluated during thermodynamic equilibrium and not traced during a phase transformation, as will be conducted for the NPs in the present study. In equilibrium, the residual hydrostatic stress in the NP core (negative Laplace pressure) is directly proportional to the surface stress and inversely proportional to the NP radius \cite{Diehm2012}. Accordingly, in the present study, the evolution of the local hydrostatic stress in NP is analyzed during amorphization.

\begin{figure}
    \centering
    \includegraphics[width=\textwidth]{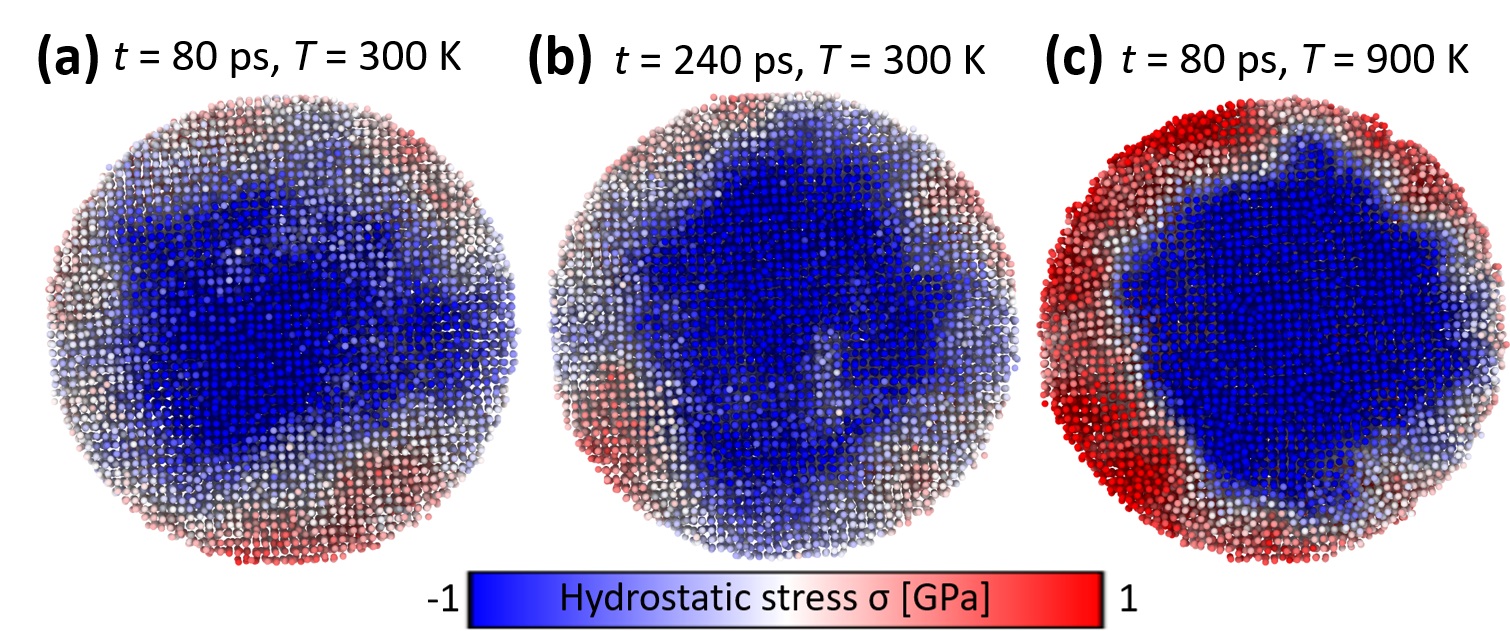}
    \caption{The hydrostatic stress $\sigma$ in the cross-section of the 10 nm NP after thermal equilibration \textbf{(a)} for 80 ps at $T=300$ K and \textbf{(b)} for 240 ps at $T=300$ K, and \textbf{(c)} for 80 ps at $T=900$ K.}
    \label{fig:hydro_300}
\end{figure}

First, the hydrostatic stress during equilibration of the 10-nm NP at $T=300$ K is traced to confirm the presence of surface stress in the initial state: see Figs. \ref{fig:hydro_300}\textbf{a,b} as pertaining to 80 ps and 240 ps of equilibration at 300 K, respectively. The atoms are colored according to the average hydrostatic stress among their neighbors in a sphere with a radius of 2 nm. The cross-section shows that the NP experiences compressive stresses in the NP core region, which are balanced by positive surface stresses (tension). The NP surface can be considered to be relaxed at the cost of straining the underlying bulk layers until a balance between compressive bulk and tensile surface stresses are reached.\cite{Diehm2012} The counteracting tension and compression regions are connected by regions of near-zero stress. The corresponding stress magnitudes are in the range of $\pm1$ GPa. Except for some small fluctuations in the stress field near the surface, this appears to be the equilibrium stress distribution in the 10 nm NP at $T=300$ K. Notably, at 300 K, the resulting stress distribution is somewhat heterogeneous (see Fig. \ref{fig:hydro_300}\textbf{b}), which is a direct result of the strong dependence of the surface stress on crystallography due to the anisotropy of the stiffness tensor (i.e. for fcc phases the 111 direction is relatively stiff) and the unfavorable high energy of non-reconstructed polar oxide surfaces.\cite{Jeurgens2000, Reichel2008b, Diehm2012} Only at a relatively high equilibration temperature of 900 K, a homogenization of the tensile and compressive stress distributions in the subsurface and core regions is reached, as aided by the $\gamma\rightarrow{am}$-\ce{Al2O3} transformation: see Fig. \ref{fig:hydro_300}\textbf{c}.

\begin{figure}
    \centering
    \includegraphics[width=\textwidth]{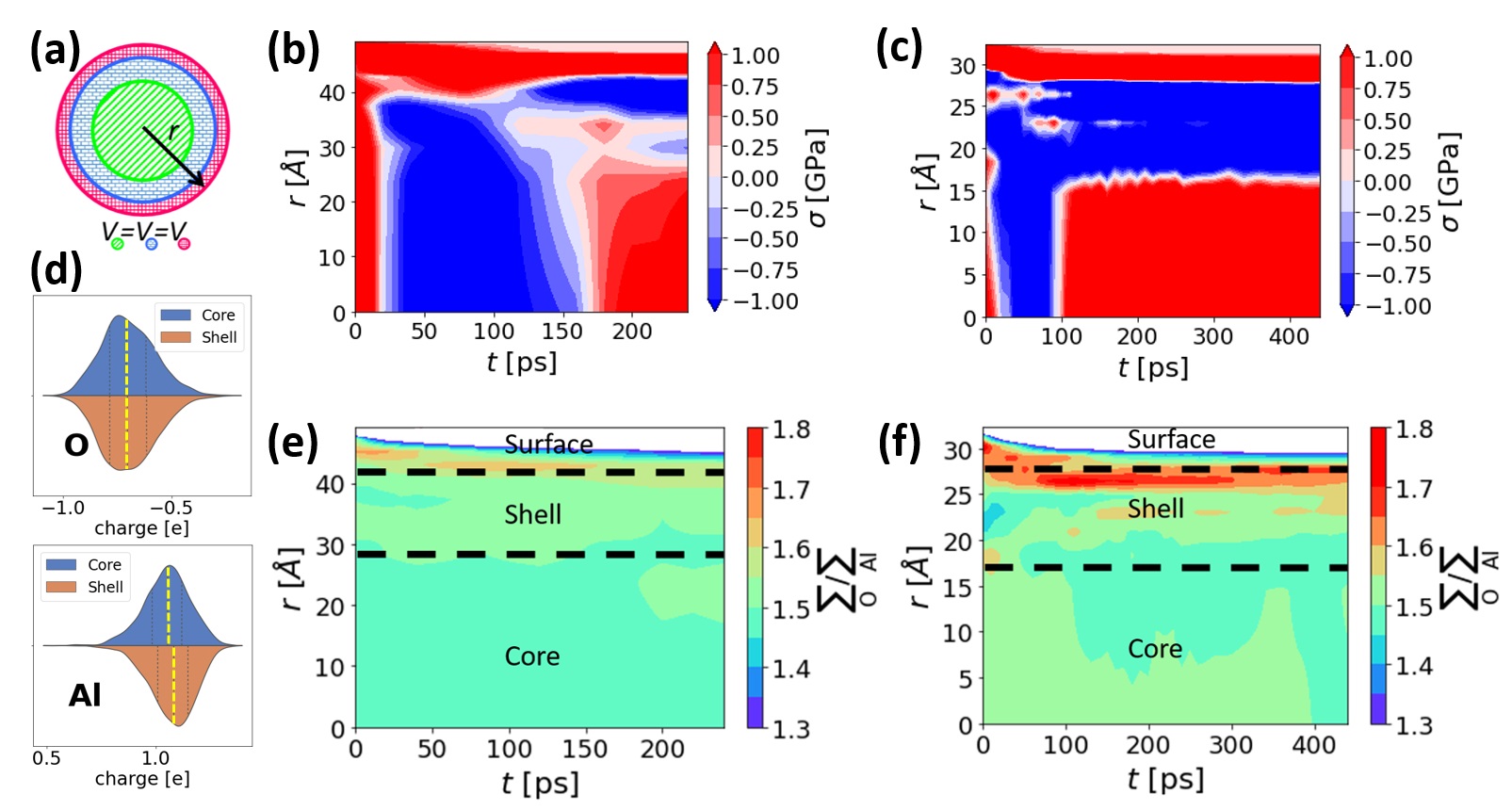}
    \caption{\textbf{(a)} A schematic of the employed spherical binning procedure to determine the average hydrostatic stress in the NP as a function of the distance, $r$, from the NP center. A total of 10 spherical bins of equal volume are defined, each of which contains approximately 1500 and 6000 atoms for the 6-nm and 10-nm NP, respectively. \textbf{(b,c)} Contour plots of the averaged hydrostatic stress, $\sigma$, as a distance from the NP center, $r$, with increasing equilibration time, $t$, \textbf{(b)} for the 10-nm NP at 900 K and \textbf{(c)} for the 6-nm NP at 900 K. Positive and negative $\sigma$-values correspond to tension and compression stresses, respectively. \textbf{(d)} Histograms of the average charge distributions (in elementary charges) of Al and O ions inside the core and the shell of the 6-nm NP, as calculated over equally sized trajectory chunks between $t=$ 160-440 ps. The yellow dashed and back dotted lines indicate the median and the first quartiles of the charge distributions, respectively. \textbf{(e,f)} Contour plots of the $\sum_{\rm{O}}/\sum_{\rm{Al}}$ atomic ratio as a distance from the NP center, $r$, with increasing equilibration time, $t$, \textbf{(e)} for the 10-nm NP at 900 K and \textbf{(f)} for the 6-nm NP at 900 K.
    \label{fig:hydrostatic_stress}}
\end{figure}

\begin{figure}
    \centering
    \includegraphics[width=\textwidth]{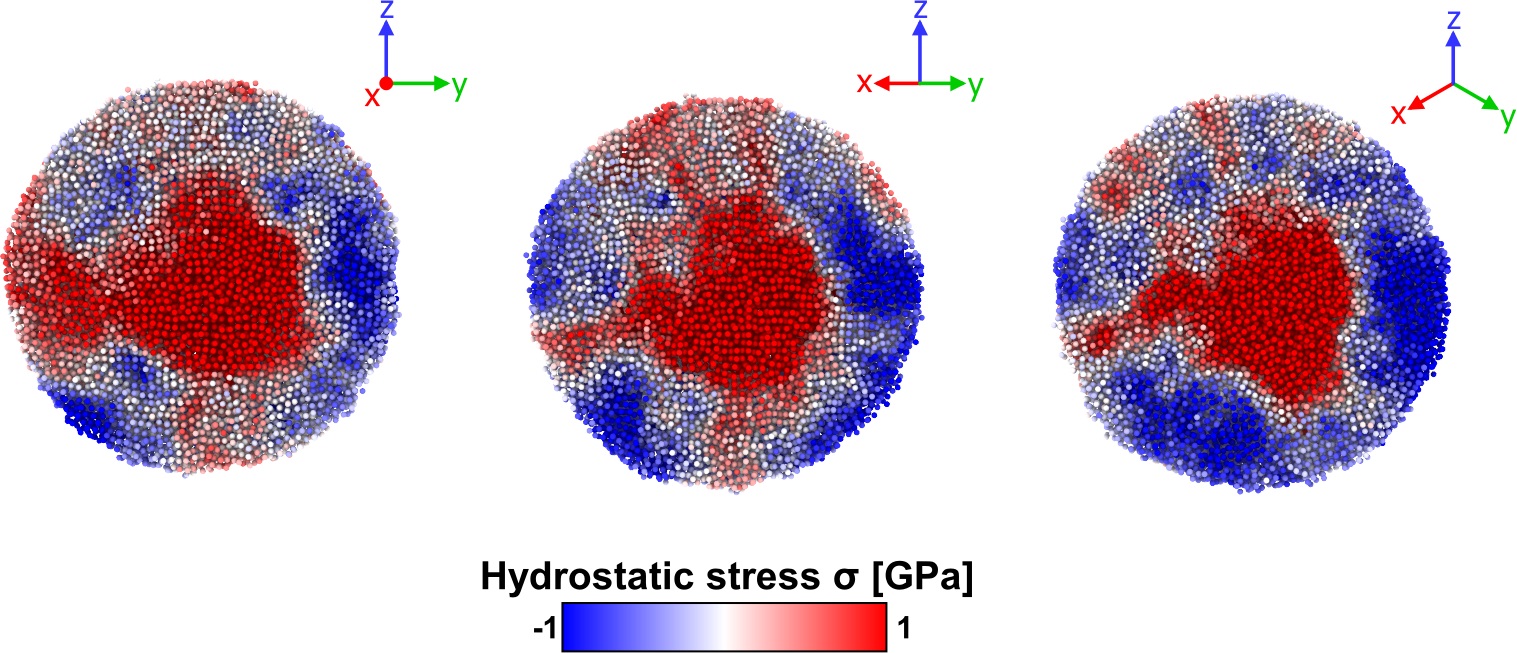}
    \caption{The final heterogeneous stress states for 10-nm NP after $t=250$ ps of equilibration at $T=900$ K. Three radial cross-sections along 100 (left), 110 (middle), and 111 (right) planes are presented. The complete evolution of hydrostatic stress over over time is given in the supplemental materials.}
    \label{fig:three views}
\end{figure}

The redistribution of the NP stresses during amorphization at 900 K can be traced by evaluating the average hydrostatic stress, $\sigma$, as a distance from the NP center, $r$, with increasing equilibration time, $t$. Fig. \ref{fig:hydrostatic_stress}\textbf{a} shows a schematic of the adopted binning procedure, which ensures equal volumes for each slice. The resulting evolutions of $\sigma$ over time for the 10-nm and 6-nm NPs are presented by the contour plots in Figs. \ref{fig:hydrostatic_stress}\textbf{b} and \ref{fig:hydrostatic_stress}\textbf{c}. At the beginning of the simulation ($t = 0$), both NPs are under bulk tension, which is an artifact of the (yet unrelaxed) free surfaces amplified by long-range Coulombic interactions. Already during the first 20 ps of equilibration, the strong capillary forces induce a shrinking of both NPs, as associated with an inversion of the stress states in the NP core from tensile to compressive, while maintaining a very high positive surface stress. A similar shrinking of \ce{MgO} NPs with increasing temperature has also been observed experimentally.\cite{Cimino1966}. The stress distribution obtained after initial NP shrinking remains about constant during the $\gamma\rightarrow{am}$-\ce{Al2O3} transformation up to about 90 ps and 170 ps for the 6-nm and 10-nm NP at 900 K, respectively. Given that the number of atoms $N_{\text{atoms}}$ in NPs scales proportionally to $r^3$, longer trajectories were only generated for the small NP. This approach enables more efficient use of computational resources for the validation of long-term trends and effects. Strikingly, once the $\gamma\rightarrow{am}$-\ce{Al2O3} transformation rate decelerates (i.e. after roughly 90 ps and 170 ps for the 6-nm and 10-nm NP, respectively; see Fig. \ref{fig:rates}\textbf{a-d}), the compressive stress state in the bulk NP becomes inverted into a tensile stress state, which is separated from the tensile surface stress region by a diffuse interface region of compressive stress with an approximately constant thickness of about 1 nm (for both the 6-nm and 10-nm NP). The existence of such a diffuse interface region between the outer surface and the NP core, in which the surface stresses are largely relaxed before reaching the bulk-like region in the NP interior, has also been reported theoretically for the atomistic modeling of crystalline and amorphous gold NPs (at 0 K)\cite{Holec2021}, as well as experimentally for the nucleation and growth of \ce{FePt} NPs.\cite{Zhou2019} The NP stress distribution after the amorphization process at 900 K can thus be described by three co-existing regions (see white dashed lines in Figs. \ref{fig:hydrostatic_stress}\textbf{(b,c)}): (\textbf{$i$}): an outer surface region with a thickness of about 1-2 atomic layers (i.e. $\sim{0.4}$ nm), characterized by a very high tensile surface stress; (\textbf{$ii$}) a diffuse interfacial region with an approximate thickness of 1 nm and a compressive stress state (further designated as shell); (\textbf{$iii$}) a NP core region with a high tensile stress state. Fig. \ref{fig:three views} illustrates the resulting heterogeneous stress distribution for three different cross-sectional cuts of a 10-nm NP (after $t=250$ ps at $T=900$ K).

The predicted \textit{tensile} stress state in the core of the equilibrated \textit{am}-\ce{Al2O3} NPs is in contradiction with the calculated \textit{compressive} stress state in the core of spherical, amorphous Au NPs at 0 K by Holec \textit{et al.} \cite{Holec2021}
However, such a presumed amorphous state of Au NPs at 0 K will be highly unstable at elevated temperatures and should thus be considered as an artificially reconstructed ("kinetically frozen") state. In contrast, the crystalline-to-amorphous transition of the $\gamma$-\ce{Al2O3} NPs at 900 K in the present study is thermally (i.e., kinetically) activated and favored by thermodynamics (cf. Fig. \ref{fig:equilibration} \textbf{a}). This is also evidenced by our model predictions that the $\gamma\rightarrow{am}$-\ce{Al2O3} transformation is faster and completed within shorter equilibration times for the smaller 6-nm NP due to its higher surface-to-volume ratio (see Fig. \ref{fig:hydrostatic_stress}\textbf{c}).

Both lattice expansion and contraction of NPs (as compared to the strain-free bulk lattice constant) have been reported, depending on the type of chemical bonding \cite{Diehm2012,Palosz2002}. Purely ionic compounds (salts) or metals, on the one hand, show lattice contraction (i.e. compression) of the NP core due to the Laplace pressure. Oxide materials, on the other hand, often exhibit lattice expansion (i.e., tension) of the NP core, as attributed to local heterogeneities in the surface charge distribution as a result of a highly defective crystal structure, which can cause a charge imbalance in the core of metal-oxide NPs with strong ionic bonding, such as alumina.\cite{Diehm2012} However, amorphous metal oxides have not yet been considered in such theoretical and experimental assessments, although they can be considered as highly disordered structures. Therefore, the charge distributions of the anions and cations in the core of the fully amorphous 6-nm $\ce{Al2O3}$ NP were compared to those in the shell region, as visualized by the respective histograms in Fig. \ref{fig:hydrostatic_stress}\textbf{d}. It follows that the average charge distribution in the core and shell regions is similar for the \ce{O} ions, whereas a noticeable difference between the core and the shell occurs for the \ce{Al} ions. A charge equilibration procedure was applied to the NPs during each equilibration step and therefore any remaining differences in the charge distribution must originate from local deviations in the stoichiometric composition, i.e. from local deviations in short-range order, as resulting in the appearance of different Al-LCS types (see Figs. \ref{fig:rates}\textbf{b,d}). Figs. \ref{fig:hydrostatic_stress}\textbf{e,f} show the corresponding contour plots of the $\sum_{\rm{O}}/\sum_{\rm{Al}}$ atomic ratio as a distance from the NP center, $r$, with increasing equilibration time, $t$ (at 900 K). Indeed, the outer surface of the $\gamma$-\ce{Al2O3} NPs is practically instantaneous enriched in Al at their outer surfaces in order to reduce the polarity and, thereby, the energy of the surface, in accordance with theory\cite{Jeurgens2000, Reichel2008b} and experiment\cite{Jeurgens2002} for \ce{Al2O3} thin films. The instantaneous reconstruction of the $\gamma$-\ce{Al2O3} NP surface creates a disbalance of charges between the outer surface and the subsurface. With increasing equilibration time (i.e. during the amorphization process), the Al-depleted subsurface region partially extends into the NP core. Consequently, after the amorphization process has ceased, the \ce{Al}-enriched outer surface is separated from the stoichiometric core region by a diffuse interfacial transition region with a slight overall Al depletion. These local deviations from the stoichiometric composition also rationalize the appearance of Al-LCS with an average coordination $> 6$ during the $\gamma\rightarrow{am}$-\ce{Al2O3} transformation of NPs (see Figs. \ref{fig:rates}\textbf{b,d}).

Hence, the model predictions show that the energetically-favored reconstruction of the initial $\gamma$-\ce{Al2O3} NP surface by Al surface segregation creates an instantaneous depletion of Al in the NP subsurface region, which gradually diffuses into the NP core. As such, a disbalance of atomic charges is created between the stoichiometric core region and the non-stoichiometric subsurface region; these separated space-charge regions result in a net attractive Coulombic force between the surface and core region of the NP, which is strong enough to oppose the Laplacian surface stress and even induce a substantial tensile stress state in the NP core (see Fig. \ref{fig:hydrostatic_stress}\textbf{b,c}). In this regard, it should be emphasized that the well-known reconstruction of polar oxide surfaces strongly depends on the environment, such as humidity (resulting in surface hydroxylation). Hence, the application of alumina NPs in gaseous or aqueous environments for e.g. catalysis will likely result in different thermal stability and/or stress states, which remains a subject of our future work. Nevertheless, the model prediction provides a valid fundamental explanation for the long-standing puzzle of why metal oxide NPs exhibit an expansion (instead of shrinkage) with decreasing size. Clearly, the above-reported phenomenon goes beyond the particular system studied here (i.e. \ce{Al2O3}), which calls for further theoretical and experimental validations of other metal-oxide NP systems.

For experimental validation of our findings, one can utilize diffraction-based techniques. However, experimental diffraction studies on measuring lattice parameters are challenging for NPs because they assume theoretically infinitely large diffraction domains. Moreover, size effects can only be revealed when probing an NP assembly with a very narrow (preferably monodisperse) size distribution. In practice, the broad size and shape distributions, as well as the inevitable aggregation of NPs in commercial oxide nanopowders have resulted in contradictory experimental findings, such as the size-dependent phase transformation sequence and respective rates for the reduction of CuO NP to Cu metal.\cite{Unutulmazsoy2022} Nevertheless, \textit{in situ} Bragg coherent diffraction imaging was able to track changes in the surface strain of NPs in relation to different reactive environments \cite{Dupraz2022}, although at much longer timescales than accessible to MD simulations. An alternative indirect and destructive way to access internal stress states in NPs is to measure their mechanical response to external deformation, which correlates with the shape and surface state of nanoparicles \cite{Sharma2018,Sharma2020}.

Possibly, more conclusive and non-invasive experimental observations might be extracted using nanophotonic approaches. Here NPs with sizes $<5$ nm are expected to comprise discrete optical resonances, which are strongly dependent on size, composition, and dielectric properties.\cite{Benz2016} Combining such NPs with Raman active molecules one can push even further the detection limit of sizes down to single atoms. Recent examples show methodologies that resolve atomic migration, subangstrom ambient motion \cite{Xomalis2020,Griffiths2021}, and infrared radiation detection in nanoscale gaps \cite{Xomalis2021}, when changing their scattering (elastic/inelastic) properties. Furthermore, our theoretical findings on NPs with different hydrostatic residual stress domains (resulting in different density and mechanical properties such as Elastic Moduli) can be experimentally investigated utilizing the acoustic oscillations of such spherical NPs. \cite{Zijlstra2008,Deacon2017} Briefly, utilizing ultrafast time-resolved detection of reflection from nanoscale objects allows resolving the evolution of stress within a few hundred of picoseconds, fitting well the timescales we found here. All the above-mentioned experimental demonstrations mainly concern studies involving crystalline metallic NPs (mainly Au); however, similar size change detection occurs in materials such as Al/\ce{Al2O3} \cite{Clark2019}, Si/\ce{SiO2} \cite{Milichko2018} and several other all-dielectrics and alloys \cite{Baranov2017,Xu2021a}. Thus, we anticipate that our theoretical findings can support the first steps of nanophotonic experimental studies on understanding atomic-scale size changes of NPs at elevated temperatures as well as explain residual stress dynamics in nanoconstructs after illumination with short laser pulses.

\section{Conclusions}
A recently developed COMB3 potential for the Al-O system was applied to investigate the thermal stability of a bulk $\gamma$-\ce{Al2O3} crystal, as well as spherical $\gamma$-\ce{Al2O3} nanoparticles (NP) with diameters of 6 nm and 10 nm, by MD simulations. The predicted defect structure of the $\gamma$-\ce{Al2O3} crystal after thermal equilibration at 900 K (i.e. about 100 K below the bulk transition temperature to $\alpha$-\ce{Al2O3}) is in excellent agreement with previously reported experimental findings. In particular, the bulk transformation from $\gamma$-\ce{Al2O3} to $\alpha$-\ce{Al2O3}, which requires a transformation of the O-sublattice from FCC to HCP, is still kinetically hindered at 900 K. However, local distortions of the FCC O-sublattice by the formation of quasi-octahedral local coordination spheres of Al (originally residing in 4-fold coordination) are thermally activated, driven by the partial covalency of the Al-O bond in \ce{Al2O3}. Such local distortions of the FCC O-sublattice in transition alumina phases have been commonly reported, but generally (naively) attributed to the formation of penta-coordinated \ce{Al} cations (based on experimental NMR). As postulated in the present work, these presumed penta-coordination spheres of Al may actually arise from local distortions of the FCC O-sublattice by the formation of quasi-octahedral Al local coordination spheres. Our simulations thus provide a general fundamental understanding of the polymorphic phase transformations of the transition alumina phases.

In contrast to the metastable $\gamma$-\ce{Al2O3} crystal, spherical $\gamma$-\ce{Al2O3} NPs with a diameter of 6 nm and 10 nm experience a thermodynamically-preferred crystalline-to-amorphous (i.e. $\gamma\rightarrow{am}$-\ce{Al2O3}) transformation at 900 K, which is faster and completed within shorter times for the smaller NP size. The amorphization process begins at the reconstructed NP surface and gradually propagates into the NP core through local structural relaxations (i.e. through collective displacements of anions and cations) resulting in the formation of $> 6$-fold local coordination spheres of Al. The resulting amorphous structure can be described by a random network of corner-sharing O-polyhedra around Al cations, which provides the well-recognized bond flexibility characteristic for amorphous oxides.\

At 300 K, the surfaces of the equilibrated 6-nm and 10-nm $\gamma$-\ce{Al2O3} NPs are reconstructed to reduce the polarity and, thereby, the energy of the surface. This surface reconstruction is realized by the fast segregation of Al from the subsurface region toward the outer surface. The initial surface reconstruction is accompanied by an NP shrinkage due to the acting Laplace pressure, which induces a compressive (negative) stress state in the NP core region (as balanced by a high positive surface stress). During the $\gamma\rightarrow{am}$-\ce{Al2O3} transformation at 900 K, the \ce{Al}-enriched outer surface and its associated \ce{Al}-depleted subsurface become separated from the stoichiometric core region by a diffuse interfacial transition region with a slight overall Al depletion. This compositional heterogeneity creates a disbalance of charges between the outer surface and the subsurface of the NP, thus inducing a net attractive Coulombic force between the surface and the core region of the NP which is strong enough to oppose the Laplacian surface stress and inducing a slight tensile stress state in the NP core.

The aforementioned model predictions thus disclose the delicate interplay between space-charge regions (due to compositional heterogeneities) and residual stresses in oxide nanosystems, which calls for improved theoretical modeling approaches. Namely, most existing theoretical frameworks are focused on crystalline and liquid systems and cannot be applied to amorphous nanomaterials, especially to those with strong ionic bonding, like metal oxides with multiple competing metastable polymorphs. With respect to the targeted applications (see Introduction), our work highlights that $\gamma$-\ce{Al2O3} NPs are not stable if processed towards higher temperatures, which has major implications not only for heterogeneous catalysis but also for the fabrication of alumina-NP-based nanocomposites by additive manufacturing routes. Our findings suggest that a complete or partial transition of $\gamma$-\ce{Al2O3} NPs to the amorphous state could affect their dissolution kinetics in liquid metals and also change their behavior as a grain nucleation agent during the solidification of liquid metals in additive manufacturing. Thus, these findings call for future investigations of chemically active surface sites in amorphized NPs (for heterogeneous catalysis), as well as of metal-oxide interface structures formed between amorphized $\gamma$-\ce{Al2O3} NPs and liquid metals (for advanced manufacturing of metal matrix composites).

\section{Acknowledgments}

This research was supported by the NCCR MARVEL, a National Centre of Competence in Research, funded by the Swiss National Science Foundation (grant number 205602). The authors acknowledge computing resources from GENCI-CINES and GENCI-TGCC (Grant 2020 - A0090912032 and Grant 2021 - A0110912032) and CSCS (project s1130). A.X. acknowledges support from the Empa internal funding call (IRC 2021).

\bibliography{References.bib}

\end{document}